\documentclass[11pt,a4paper,leqno]{article}

\usepackage{amsmath,amssymb,amsthm}
\usepackage{epsfig}

\usepackage{amsmath}
\usepackage{amssymb}
\usepackage{makeidx}
\usepackage{graphicx}
\usepackage{color}

\newcommand{\hide}[1]{}

\newcommand{\up}[1]{ ^{(#1)}}
\newcommand{\inv}[1]{ ^{-#1}}

\newcommand{\myvec}[1]{\vec{#1}}

\newcommand{\vc}{{\myvec{c}}}

\newcommand{\tpsi}{\tilde{\psi}}

\newcommand{\mymat}[1]{\widehat{#1}}

\newcommand{\mD}{\mymat{D}}

\newcommand{\mS}{\mymat{S}}

\newcommand{\RR}{ {I\!\!R} }
\newcommand{\CC}{ {I\!\!\!\!C} }
\newcommand{\D}{\mathcal{D}} 
\renewcommand{\H}{\mathcal{H}} 

\renewcommand{\r}{\rangle}
\renewcommand{\l}{\langle}

\newcommand{\ddt}{\frac{d}{dt}}

\newcommand{\ddx}{\partial_x}

\newcommand{\om}{\omega}
\newcommand{\si}{\sigma}

\newcommand{\al}{\alpha}
\newcommand{\la}{\lambda}

\renewcommand{\th}{\theta}
\newcommand{\Th}{\Theta}

\renewcommand{\bar}{\begin{array}{ll}}
\newcommand{\ear}{\end{array}}
\newcommand{\bma}{\begin{pmatrix}}
\newcommand{\ema}{\end{pmatrix}}
\newcommand{\beq}{\begin{equation*}}
\newcommand{\eeq}{\end{equation*}}
\newcommand{\bel}[1]{\begin{equation}\label{eq:#1}}
\newcommand{\eel}{\end{equation}}
\newcommand{\bea}{\begin{eqnarray*}}
\newcommand{\eea}{\end{eqnarray*}}

\newcounter{lecture}

\newcommand{\Ef}{\mathcal{E}}

\theoremstyle{remark}

\def\l{\langle}
\def\r{\rangle}

\theoremstyle{plain}

\newcommand{\zr}{ {I\!\!R} }
\newcommand{\R}{ {I\!\!R} }

\renewcommand{\d}{\partial}

\renewcommand{\d}{\partial}

\renewcommand{\ddt}{\frac{\d}{\d t}}

\numberwithin{equation}{section}

\title{On the non-equivalence of perfectly matched layers and exterior complex scaling}

\author{A. Scrinzi$^a$, H. P. Stimming$^b$, N. J .Mauser$^b$
\\
$^a$Ludwig Maximilians Universit{\"at}, \\
Theresienstrasse 37, 80333 Munich, Germany;
\\
$^b$Wolfgang Pauli Institut $\mathrm{^c\!/\!_o}$ Department of Mathematics,
\\
Universit{\"a}t Wien, Nordbergstrasse 15, A1090 Vienna, Austria.}

\begin{document}
\maketitle

\begin{abstract}
The perfectly matched layers (PML) and exterior complex 
scaling (ECS) methods for absorbing boundary conditions  
are analyzed using spectral decomposition.
Both methods are derived through analytical continuations from unitary to contractive
transformations. We find that the methods are mathematically and numerically distinct: ECS is 
 complex {\em stretching} that {\em rotates} the operator's spectrum into the complex plane, 
whereas PML is a complex {\em gauge transform} which {\em shifts} the spectrum.
Consequently, the schemes differ in their time-stability. Numerical examples are given.
\end{abstract}



\section{Introduction}
An important problem in numerical simulations of time dependent 
equations is the need  to restrict  infinite domains to finite computational domains
for purposes of approximation.
A simple cutoff of the domain by trivial boundary conditions
(e.g. imposing zero Dirichlet boundary conditions, or periodic continuation) 
leads to  errors at the boundary of the (artificial) domain of simulation 
as soon as the solution is not small enough anymore near the boundary. 
Much work has been dedicated to the question of how to either  remove these errors 
totally or to sufficiently suppress them.
A popular method to prescribe artificial absorbing boundary conditions 
is the perfectly matched layer method (PML).
A slightly older method uses 
a very similar, but nonetheless distinct approach
called exterior complex scaling (ECS). 
This method has mainly been applied 
to dispersive problems, like, e.g., the Schr{\"o}dinger equation.

The ECS was proposed by B.~Simon \cite{simon79:complex-scaling} as an extension of the 
complex scaling method first established
by Balslev and Combes \cite{Balslev1971} for
Schr\"odinger operators. 
The PML approach initially was introduced by Berenger \cite{berenger94:pml} for
application to Maxwell equations or wave equations, but later also was
applied to Schr{\"o}dinger type equations.

There exist several other methods to treat the domain cutoff problem, for example
transparent boundary conditions,
complex absorbing potentials (optical potentials), 
or time dependent phase space filters,
which are not subject of this work.
An overview of  most of these methods can be found in \cite{antoine08:absorption}.

A large body of mathematical physics literature exists on ECS (many references 
can be found, e.g., in \cite{brandas89_resonances}). Early computational applications
were for the computation of decaying states in quantum chemistry (see, e.g., \cite{reinhardt82,scrinzi:jcp1993}).
In \cite{scrinzi10:irecs} the numerical realization of ECS was
improved by using infinite range basis elements in the boundary region
and thus considerably increasing the efficiency of the method.

PML was applied to Maxwell equations by Chew and Wheedon, e.g., in \cite{chew94:pml}.
A recent study on PML  for general hyperbolic problems with (algorithmic) stability 
is found in \cite{appelo06:pml-hyperbolic}.
We mention also a work by Lions et.al. on a stabilized PML method \cite{LionsMetral2002}.
Concerning the application of PML to the Schr{\"o}dinger case, the first 
study was  \cite{collino97:pml}. In this work, the PML method when
applied to the  Schr{\"o}dinger equation was actually modified to be conceptually
equivalent to the ECS method. As far as we are aware, the same is true for other works
on PML for Schr{\"o}dinger type equations. 
Recent contributions to this subject are  
\cite{Zheng2007,Dohnal2009,nissen11:pml-schrodinger}.
Loh et.al. studied the failure of the PML method for wave guides with 
backward wave structures \cite{loh09:pml-failure}. 

ECS and PML share the general idea of using analytic continuation of the equation outside 
a finite spatial domain. However, in their respective original formulations, the approaches 
differ qualitatively. This is already evident from the absence of ECS methods for wave equations 
and from the comparatively poor performance of PML for the TDSE. In the present work we analyze both 
methods in a common analytic approach using spectral analysis. This clearly exposes the differences between them
and explains their different stability behavior for Schr\"odinger-like and wave-like equations,
respectively. The distinct spectral characteristics have immediate
bearings on the numerical implementation. In brief, ECS 
{\em rotates} the continuous spectrum into the complex plane, while
PML {\em shifts} the spectrum. In the spatial domain this corresponds to stretching of
the coordinates and a complex gauge transform, respectively. PML is applicable for operators like $-i\ddx$ 
whose spectrum covers all $\RR$, while ECS works well on Schr\"odinger operators with continuous spectrum 
on $\RR_+$. We could not formulate a version of ECS that would work with $-i\ddx$,
while we could obtain highly accurate results for the TDSE with both, ECS and PML. With the Schr\"odinger
equation, we found PML to be less accurate than ECS, where full computational precision could be reached.

In the following, after specifying the wave and Schr\"odinger equations,
we present in detail the spectral analysis of ECS. The analogous discussion is presented
for PML, before we give numerical examples. We conclude with pointing to the potential 
of the approach for designing absorbing boundary methods for general linear wave equations. 

\section{Model equations}
The goal is to provide absorbing boundaries for  time-dependent 
evolution equations. The class of problems usually treated by these methods 
includes, both, dispersive and hyperbolic equations, and in general problems which allow  wave-type solutions. 
For the purposes of this discussion, we consider the
time-dependent Schr\"odinger
equation and a wave equation typical model problems from this class, which can both be formulated
as a first order propagation equation

\begin{equation}
\label{evol_D}
i \frac{\d}{ \d t} \phi = D \phi
\end{equation}
with a self-adjoint operator $D$ acting on the space variable.

The scalar wave equation in reads
\bel{wave}
i\frac{\d}{ \d t} \phi = - i \nabla  \phi.
\eel
The time-dependent Schr\"odinger equations reads
\[
i \frac{\d}{ \d t} \phi[t](x) = [ -\frac 1 2 \Delta +V(x)] \phi[t](x).
\]
where $V(x)$ is a potential term. ECS is rigorously applicable for non-zero $V(x)$,
but for the sake of simplicity in the present discussion, 
we restrict ourselves to $V(x)\equiv0$.

\section{Spectral analysis of ECS}

When solving Eq.~(\ref{evol_D}) one is usually interested in the solution $\phi[t](x)$ 
only on a finite domain $x\in F \subset \R$. Outside $F$ the solution can be damped or otherwise removed.
A basic assumption of ECS, and equally of PML, is that asymptotically the solution takes the 
form 
\[
\phi[t](x) \stackrel{|x|\to\infty}\sim \int_0^\infty d\om \tpsi(\om)\exp[i|k(\om)||x|], 
\]
i.e.\ it has outgoing asymptotics. The functional dependence of the wave vector
$k(\om)$ on the frequency $\om$ determines the system's dispersion. The idea is to convert the unlimited
spatial extension of an outgoing solution to exponentially decaying form by 
continuing it into the positive complex plane for $x\not\in F$. The numerical approximation of the
exponentially decaying solution requires much smaller grids than the full solution.
Let us label the analytically continued solution by a complex number $\eta$. 
For the new equation with exponentially decaying asymptotics 
\begin{equation}\label{eq:ecs-equation}
i \frac{\d}{ \d t} \psi_\eta[t](x) = D_\eta \psi_\eta[t](x),
\end{equation}
one wants the following property to hold:
\begin{equation}\label{eq:agreement}
\phi[t](x)=\psi_\eta[t](x)\quad{\rm for}\quad x\in F, \forall t.
\end{equation}
For this, one constructs the transformation such that it is unitary for real values of $\eta$
and (\ref{eq:agreement}) holds. For complex $\eta$ one can obtain an asymptotically decaying solution
and use analyticity arguments to show that (\ref{eq:agreement}) holds.

A second requirement crucial for numerical use of (\ref{eq:ecs-equation}) is stability: 
for propagation towards positive times, the spectrum of the discrete representation 
of $D_\eta$ must be in the lower half complex plane.
Spectral components with positive imaginary parts will lead to exponential growth of the
$L^2$-norm of $\psi_\eta$ and to numerical breakdown of the time-propagation, 
irrespective of the particular propagation scheme used. 

The necessary spectral properties under these transformations 
have been proven for a wide class of Schr\"odinger operators that includes
few-body Coulomb systems and a large class of particles interacting by short range potentials (for a 
review see, e.g.,\cite{reed-simon82}). For those ``dilation analytic'' Schr\"odinger operators, $D_\eta$ can be
defined as an operator on the position space $\H_x=L^2(dx,\RR)$ 
with spectrum in the lower complex half-plane. Note that the 
numerical scheme must be set up such that also the discretized operators 
have no spectrum in the upper complex plane.
  
Earlier applications of complex scaling to time-dependent problems
\cite{scrinzi:pra2000,saenz02:H2-1} were using $F=\emptyset$, 
i.e.\ no directly usable information on $\phi[t](x)$ was obtained
and only matrix elements could be computed using analytical continuation. 
In recent literature, applications of ECS were reported, where, however, 
severe numerical shortcomings were observed \cite{He2007,tao09:complex_scaling}. 
The difficulties were overcome in ``infinite range ECS'' \cite{scrinzi10:irecs}, 
where results of machine precision accuracy were obtained employing as few 
as 10 grid points for absorption. The underlying mathematical rational is laid out now. 

Considering that the operator $D$ is self-adjoint on $L^2(dx,\zr)$ with domain $\D(D)$ and spectrum $\si(D)\subset \zr$ 
we can spectrally decompose the solutions as 
\[
\tpsi[t](\om)=(W\inv1\psi[t])(\om)=\int dx \overline{w(x,\om)}\psi[t](x),\quad\om\in\si(D).
\]
The integral kernel of the unitary map $W: \H_\om:=L^2\left(d\mu(\om),\si(D)\right)\to\H_x:=L^2(dx,\RR)$ defines
the spectral eigenfunctions $w(x,\om)$. We can write
\begin{eqnarray}
(D\psi)(x) 
&=& \int_{\si(D)} d\mu(\om) w(x,\om) \om \int dx' \overline{w(x,\om)} \psi(x')  
\\
&=& \int_{\si(D)} d\om \rho(\om) w(x,\om) \om \int dx' \overline{w(x,\om)} \psi(x').  
\end{eqnarray}
Here $d\mu(\om)$ denotes the spectral measure.
In the last line we have, for notational simplicity, assumed that $D$ has a purely
absolutely continuous spectrum, which allows us to extract the ``density of states'' $\rho(\om)\geq0$ 
as an absolutely continuous function. For now,
we disregard a possible multiplicity of the spectrum.
The spectral representation of $D$ is briefly written as
\beq
D= W\hat{\om} W\inv1,
\eeq
where $\hat{\om}$ denotes multiplication by $\om$ in $\H_\om$.

To arrive at the scaled equation (\ref{eq:ecs-equation}) one takes the following steps:
\begin{itemize}
\item[(a)] 
Introduce unitary scaling $U_\la$ that leaves the 
solutions invariant on $F$ for a range of {\em real} values of $\la$.
With this, define a self-adjoint (real) scaled operator $D_\la:=U_\la D U_\la\inv1$.
Ideally, show analyticity properties  
of the $D_\la$ w.r.t.\ $\la$.
\item[(b)] Solve Eq.~(\ref{eq:ecs-equation}) in the spectral domain
and show that $\psi_\la[t](x)$ is an analytic function of $\la$.  
\item[(c)] 
Analytically continue to complex values $\eta=\la+i\th$ and conclude that
$\psi_\eta[t](x), x\in F$ does not depend on $\eta$.
\item[(d)] Show that $\psi_\eta[t](x)$ does nowhere grow exponentially in time,
i.e. the analytically continued operator $D_\eta$ has no eigenvalues in the
upper complex half-plane.
\end{itemize}

\subsection{ Real scaling by $U_\la$}

Exterior real scaling is the transformation
\begin{equation}\label{eq:real-scaling}
(U_\la\psi)(x) = e^{\la \Th_F(|x|)/2}\psi(z_\la(x)),\quad z_\la(x):=\frac{x}{|x|}\int_0^{|x|} dr  e^{\la \Th(r)}
\end{equation}
with
\beq
\Th_F(|x|)=0\quad{\rm for}\quad x\in F,\quad \Th_F(|x|)>0\quad{\rm else}.
\eeq
$U_\la$ is manifestly unitary for all $\la\in\RR$. The exact choice of $\Th$ is unessential for the 
present discussion but impacts on discretization. 
Note that $\Th$ can be discontinuous.
We will use the discontinuous function $\Th(|x|)=1$ for $x\not\in F$, which
leads to analytically simple and numerically efficient scaling. 

With discontinuous scaling, one needs to worry about the correct definition of derivatives.
It is easy to show that on functions $\chi\in \D(D_\la)$ with the scaled 
domain $\D(D_\la)=U_\la\D(D)$, $D_\la$ is self-adjoint. It has a spectral representation
\begin{eqnarray}
\nonumber
D_\la &=& W_\la \hat{\om} W_\la\inv1
\\ &=& \label{eq:dlambda}
\int_{\si(D_\la)} d\om \rho_\la(\om) w_\la(x,\om) \om \int dx' \overline{w_\la(x,\om)} \psi(x'),  
\end{eqnarray}
with the scaled spectral eigenfunctions
\begin{equation}\label{eq:wspectral}
w_\la(x,\om): = (U_\la w)(x,\om).
\end{equation}
Unitarity of $U_\la$ ensures that $\si(D_\la)=\si(D)$. Note, however, that
functions $\chi_\la\in\D(D_\la)$ and also their derivatives are discontinuous at the borders $x_0\in\partial F$ of $F$
in the form
\begin{eqnarray}\label{eq:discontinuity}
\chi_{\la}(x_0+0)&=&e^{\la/2}\chi_\la(x_0-0)\\
\chi'_{\la}(x_0+0)&=&e^{3\la/2}\chi'_\la(x_0-0).
\end{eqnarray}
The functions
$w_\la(x,\om)=(U_\la w)(x,\om)$ are orthonormal in the sense
\beq
\int_\RR dx\, \overline{w_\la(x,\om)}w_\la(x,\om')=\rho(\om)\delta(\om-\om').
\eeq
The spectral representation is defined by a unitary
transformation to an $L^2$-space of functions on the spectrum of the operator, 
where the operator appears as multiplication by $\om$. Therefore 
the spectral density $\rho_\la(\om)\neq\rho(\om)$ 
must be considered as $\la$-dependent, when we use the $w_\la(x,\om)$ defined in (\ref{eq:wspectral}). 
The spectral resolution of identity is
\begin{equation}\label{eq:resolution}
\chi(x)=\int_{\si(D)} d\om \rho_\la(\om) w_\la(x,\om) \int_\RR dx' \overline{w_\la(x',\om)}\chi(x').
\end{equation}
Let us illustrate the $\la$-dependence of $\rho_\la$ in the simple case of global scaling $F=\emptyset$
for the wave-equation with $D=-i\ddx$. For a scaled spectral eigenfunction it acts as
\beq
-i\ddx U_\la \frac{1}{\sqrt{2\pi}}e^{i\om x} 
= -i\ddx \frac{1}{\sqrt{2\pi}}e^{i\om e^\la x}
=\frac{e^{\la}}{\sqrt{2\pi}}\om e^{i\om x}.
\eeq
We see that the set of spectral eigenfunctions and overall spectrum remain the same, but the association 
of a given $\om$ with individual spectral functions $w_\la(x,\om)$ changes, in this case by the factor 
$e^{\la}$. This factor must be compensated
by a change in the density of states. For this particular example $\rho_\la(\om)=e^{-\la}$. 
For $F=\emptyset$ we have $\D(D_\la)=\D(D)$ and we may just as well re-define
the integration variable $\om\to\om'=e^\la\om$ and find
$U_\la\ddx U_\la\inv1=e^{-\la}\ddx$, a rather obvious result in this case.

As it turns out, the modification of the spectral density for the operators
$-i\ddx$ and $-\ddx^2$ depends only on the asymptotic behavior of the solution and it is 
independent of any finite $F$. This also holds when adding a ``dilation analytic''
potential to the differential operators. In fact, the $D_\la$ can 
be considered as an operator-valued analytic function. We refer the reader to 
\cite{reed-simon82} for a detailed discussion.
Accepting this assertion, we can continue our formal discussion in the spectral domain.

For solving (\ref{eq:ecs-equation})
we go into the spectral representation by applying the resolution of identity (\ref{eq:resolution})
\bel{om-equation}
i\ddt \tpsi_\la[t](\om) = \rho_\la(\om)\om \tpsi_\la[t](\om).
\eel
In $\H_\om$ the solution is
\bel{om-solution}
 \tpsi_\la[t](\om) = e^{-i\rho_\la(\om)\om t} \tpsi_\la[0](\om),\quad \tpsi_\la[t](\om)=(W_\la\inv1\psi_\la[t])(\om)
\eel
and in $\H_x$ it is
\bel{x-solution}
\psi_\la[t]=\int d\om \rho_\la(\om) w_\la(x,\om) e^{-i\rho_\la(\om)\om t} \tpsi_\la[0](\om).
\eel
For any unitary $U_\la$, it trivially holds that
\beq 
\psi_\la[t](x)=U_\la\psi[t](x).
\eeq
Equality is to be understood in the $L^2$-sense on sets of measure $>0$, not point-wise.
As our exterior scaling $U_\la$ changes functions only outside $F$ we have
\bel{invariant}
\psi_\la[t](x)=\psi[t](x)\quad{\rm for}\quad x\in F, \forall\, t.
\eel

\subsection{Analytic continuation}

We analytically continue the integrand on the r.h.s.\ of Eq.~(\ref{eq:x-solution})
 to complex values of $\la$. If the initial state $\psi[0](x)$ has its support
in $F$, its spectral components are independent of $\la$
\beq
\tpsi_\la[0](\om) = \int_F dx \overline{w_\la(x,\om)}\psi[0](x) 
= \int_F dx \overline{w(x,\om)}\psi[0](x)=\tpsi[0](\om).
\eeq
$\rho_\la(\om)$ is $e\inv{\la}$ and $e\inv{2\la}$ for $D$ asymptotically $\sim-i\ddx$ and $\sim-\ddx^2$,
respectively. Finally, the spectral eigenfunctions are
\beq
w_\la(x,\om)\sim \exp[ik(\om)e^\la \Th(x) x],
\eeq
with $k(\om)=\om$ for $-i\ddx$ and $k(\om)=\pm\sqrt{\om}$ for $-\ddx^2$.

The spectral representation (\ref{eq:x-solution}) then reads, after analytical continuation
 $\la\to i\th$ ($\th \in \zr$)
\bel{cs-solution}
\psi_\th[t](x)=\int d\om e^{-in\th} w_\th(x,\om) e^{-i\cos(n\th)\om t}e^{-\sin(n\th)\om t} \tpsi[0](\om).
\eel
with $n=1,2$ for asymptotic behavior $\sim(-i\ddx)^n$.

The complex scaled solution $\psi_\th[t](x)$ is analytic in 
$\th$, as it is an integral of analytic functions. From analyticity
and independence of $\psi_\la[t](x)$ for real $\la$ it follows that 
the solution agrees with the unscaled solution on $F$, i.e.\ Eq.~(\ref{eq:agreement}) is satisfied.

Finally we examine the spatial and temporal behavior of the scaled solution.
If in the unscaled solution $\psi_{\th=0}[t]$ has outgoing behavior, then, by analyticity, 
the scaled solution $\psi_{\th>0}[t]$ at fixed time $[t]$ decays exponentially with $|x|$.
The operator $-\ddx^2$ has strictly positive spectral values $\om$ and
therefore, for positive $\sin(2\th)$, the solution is non-growing in time.

However,  we see that for $-i\ddx$ the solution will 
contain components that grow in time: its spectrum covers positive 
as well as negative $\om$ and therefore $\sin(n\th)\om$ cannot 
be positive over the complete spectrum for constant $\th$. This cannot be 
controlled in ECS in the present form.
A restriction of $\tpsi[0](\om)$ to one spectral half-axis is not desirable, as it excludes initial states
that are spatially confined to $F$. Even more importantly, in a computational scheme 
formulated in $\H_x$ rather that $\H_\om$, 
numerical noise will always generate components throughout the spectrum, which 
cause fatal numerical instability. Below we discuss, how 
PML avoids this problem by a different analytical continuation scheme.

In principle, similar reasoning as for temporal instability might be used for the spatial buildup
of undesired {\em ingoing} components, i.e. for the spatial growth towards grid boundaries. However,
in this case the spatial discretization itself, e.g. by using a square-integrable basis set, suppresses 
any spatially growing components. The fact that we are allowed to simply suppress ingoing components
must be derived from {\it a priori} knowledge of our solution. This is usually either exactly known
or a physically motivated assumption. 

Also, note that the suppression of ingoing waves
is only {\em asymptotic}. From any domain outside $F$, wave amplitude can return 
into $F$. In fact, clear numerical 
evidence was found in \cite{scrinzi10:irecs} that this happens without compromising the solution on $F$.
Absorption in the sense of loss of information is only caused by finite numerical accuracy
in representing the exponential tail of the solution.

\section{Spectral analysis of  PML} 
PML is often introduced in the same way as ECS. Then, in 
a final step, each spectral function $w(x,\om)$ is scaled separately by an $\om$-dependent factor.
The procedure is motivated as to create a uniform exponential decay for all wave vectors, rather
than the $k(\om)$-dependent damping of ECS. For scaling of $-i\ddx$ one replaces
in (\ref{eq:real-scaling})
$e^{\la\Th_F/2} \to 1+\la/\om\Th_F$.
A space-dependent spectral shift by $\la$ results:
\beq
U_{\la}\up{PML}e^{i\om x}=e^{i\om(1+\la/\om\Th_F(x)) x} =e^{i(\om+\la\Th_F(x)) x}.
\eeq
Indeed, the intended effect is achieved: for $\la=i\th$ all spectral functions decay
equally as $e^{-\th x}$. Depending on the numerical implementation, 
this may or may not facilitate numerical approximation.
Its crucial effect, however, is to make the scheme temporally stable by the
$\om$-dependent sign change: negative frequency components are stretched with the opposite sign
of the positive frequency components. When continuing the solution into the complex
plane, both, positive and negative frequency parts of the spectrum become 
exponentially damped simultaneously.
For functions $\phi\in\H_x$, $U_\la\up{PML}$ is a (local) gauge transform:
\beq
(U_{\la}\up{PML} \phi)(x)= e^{i\la \Theta_F(x)x}\phi(x).
\eeq
Obviously, $U_\la\up{PML}$ is unitary for $\la\in\RR$.  Let us use $D=-i\ddx$. 
On functions $\chi\in\D(D_\la\up{PML})=U_\la\up{PML}\D(D)$,
the PML derivative is 
\beq
D_\la\up{PML}\chi = e^{i\la\Theta_f(x)}De^{-i\la\Theta_f(x)}\chi = [D-\la\Theta_f(x)]\chi.
\eeq
By choosing $\chi\in\D(D_\la)$, we have made sure not to
get $\delta$-like contributions at $\partial F$. 

Now we can use the same general reasoning as in the case of exterior complex scaling to conjecture
that on $F$ the solutions of
\beq
i\ddt \psi_\la[t] = D_\la\up{PML} \psi_\la[t]=[D-\la\Theta_f(x)] \psi_\la[t]
\eeq
agree with $\phi[t](x)$, Eq.~(\ref{eq:agreement}).
Invoking analyticity arguments, this can be extended to complex values $\la\to i\th$,
for which the solutions $\psi_{i\th}[t]$ will be exponentially decaying outside $F$.
As the absolutely continuous spectrum is determined by the asymptotic behavior,
all frequencies $\om$ of the time-evolution experience a uniform spectral shift to into
the lower half plane to $\om-i\th$. 

\subsection{Frequency dependent PML for the Schr\"odinger equation}

As mentioned in the introduction, a large amount of literature exists
on the topic of scaling-based absorbing layers for time dependent Schr\"odinger equations.
The first work on the subject was done by Collino \cite{collino97:pml}, where 
the TDSE appears in the context of modeling of  waves as a paraxial approximation
of a higher dimensional equation. The frequency used in construction of the PML  
is actually a carrier frequency in the superseding high-dimensional model, which in the
paraxial TDSE equation appears only as a constant. So for the TDSE, the PML
used by Collino did not use frequency-dependent rescaling and was actually
equivalent to the ECS method. (However the rescaling was not formulated to be
unitary, consisting instead only of a coordinate transform.)
Other authors on the subject are Levy \cite{Levy2001:PML}, Farrell and Leonhardt
\cite{FarrellLeonhardt2005} and Zheng \cite{Zheng2007}, who all followed the 
same approach using a method equivalent to ECS ($\om$-independent rescaling). 
In \cite{dohnal07:pml}, the coupled mode equations are studied, which are a 
combination of the two model cases considered here, and 
combine hyperbolic operators in one
space direction and a Schr\"odinger-like operator in the perpendicular direction.
The PML method for these equations is $\om$-dependent. 
In \cite{Dohnal2009}, a slightly generalized TDSE is studied which 
includes mixed derivatives of second order. $\om$-independent rescaling is used.
The recent work of Nissen et.al. \cite{nissen11:pml-schrodinger}
uses a rescaling of both variable and wave function which is not unitary, but
instead is used  to obtain a formulation of the 
scaled operator which avoids first order derivative terms. 
Here, again, rescaling is not depending on $\om$, which makes the method of ECS type.

We will now 
investigate the behavior of $\om$-dependent PML for the TDSE.
A local gauge transform $U_\la\up{PML}$ generates
\beq
D_\la=(-\ddx^2)_\la = -\ddx^2 -2i\la\Th_F(x)\ddx + \Th_F(x)\la^2
\eeq
with the spectral values
\beq
\om\to (\pm\sqrt{\om}+\la)^2 = \om \pm 2\sqrt{\om}\la + \la^2.
\eeq
Analytic continuation to complex values $\la\to\eta=\la e^{i\al}$ invariably introduces positive 
as well as negative imaginary parts in the spectrum.
However, we may numerically suppress ingoing waves $k(\om)x<0$:
by making the gauge transform dependent on the sign of $x$
\beq
\Th(x):=\left\{\begin{array}{rl}
 0&{\rm for}\quad x\in F \\
 1&{\rm for}\quad 0<x\not\in F\\
-1&{\rm for}\quad 0>x\not\in F.
\end{array}\right.
\eeq
With $0<\la$ and complex phase $0<\al<\pi/2$, all outgoing spectral components 
decay exponentially in space. In addition, they are associated with 
spectral values in the lower complex half plane and decay exponentially in
time. Ingoing components, which would be exponentially
growing in space and time, are assumed not to occur in the solution.
They are excluded from the discrete problem by choosing square-integrable 
representation.
If the complex argument is chosen at $\al=\pi/2$, the transformation maps
$\om=0\to -\la^2$, back to the negative real axis, and time-decay of 
small-$\om$ components becomes inefficient.
Indeed, with complex arguments $\al$ approaching $\pi/2$, 
the PML for the TDSE starts to break down
(see numerical examples below).

We cannot, at this point, provide mathematical proof for this formal reasoning.
The theory of dilation-analytic operators is inapplicable here.
However, this does not mean that the absorption scheme fails. Even more severe 
mathematical questions 
arise in ECS for the Stark-effect where a potential of the form $x\Ef$ is added to the Hamiltonian.
Likewise, adding time-dependent dipole interactions $iA(t)\ddx$ as in \cite{scrinzi10:irecs} extends
the spectrum to all $\RR$. Yet, in both cases, ECS 
delivers verifiably accurate results (see, e.g., \cite{saenz02:H2-1,scrinzi98}). 

\section{Numerical evidence}

Without further investigating the mathematical arguments, we now show that 
numerical experiments confirm the above formal expectations.
For a detailed numerical study of ECS we refer to a series of recent publications
on absorption by ECS in one and three dimensions \cite{scrinzi10:irecs,tao12:ecs-spectra,scrinzi12:tsurff}. 
ECS was found to provide 
perfect (machine precision) absorption for 1d and 3d problems, even 
when the Schr\"odinger operator contains an explicitly time-dependent part.

All numerical calculations are performed using a high order finite element
spatial discretization. As the scheme relies on analyticity, when using our 
discontinuous $\Th_F$, it is important to correctly implement the discontinuity 
at $\partial F$. 
One might think of starting from basis functions $\phi_i\in\D(D)$ and use
$\phi_i\up{\la}(x)=(U_\la\phi_i)(x)$. However, in that case, for real $\la$ 
all matrix elements of the transformed equation would be trivially 
identical to the matrix elements of the original equation with the untransformed 
basis and this property would carry over to complex $\eta$. 
The correct strategy is to discretize the transformed domain $\D(D_\eta)$, where we can 
take advantage of the exponentially damped character of the solution. 
One possibility to build the discontinuity (\ref{eq:discontinuity}) into the basis 
is by using $\chi_i\up{\la}(x)=e^{-\la/2\Th_F(x)}\chi_i(x)$, where the $\chi_i$ are everywhere
differentiably 
continuous. The discontinuity of the first derivative does not need to be 
imposed, if we compute matrix elements by partial integration as 
\beq
\left(\mD_\la\right)_{ij}:=\l \chi_i\up{\la} | \left(-\ddx^2\right)_\la |  \chi_j\up{\la}\r
=\int_F dx \overline{\chi'_i(x)}\chi'_j(x) 
+ e^{-\la}\int_{\RR\backslash F} \!\!\!dx\overline{\chi'_i(x)}\chi'_j(x).
\eeq 
The $e^{\la}$ from the discontinuity partially cancels with $e^{-2\la}$
from the scaled second derivative leaving the overall factor $e^{-\la}$ 
in front of the second summand. The $\la$-dependence of the basis here also
affects the overlap matrix
\beq
\left(\mS_\la\right)_{ij}=\int_F dx\, \overline{\chi_i(x)}\chi_j(x) 
+e^{\la}\int_{\RR\backslash F} dx\, \overline{\chi_i(x)}\chi_j(x).
\eeq 
Continuation $\la\to\eta\in\CC$ leads to a non-hermitian overlap matrix. 
For real $\chi_i$, $\mS_\eta$ is complex symmetric:
$\left(\mS_\eta\right)_{ij}=\left(\mS_\eta\right)_{ji}$.

In PML, the basis for real $\la$ is multiplied by a $x$-dependent   
phase, which leaves the overlap matrix elements unaffected for all real $\la$,
and no complex continuation of the overlap matrix is needed: $\mS\up{PML}_\eta\equiv\mS$.
This is a relevant technical difference between ECS and PML.

One finally solves the discretized system
\beq
i\ddt \vc = \mS_\eta\inv1 \mD_\eta \vc
\eeq 
by some standard ODE solver. In the present work we chose 
a step-size controlled classical 4th order Runge-Kutta.

\subsection{Spatial discretization, initial state, propagation time}
No effort was made to optimize the spatial discretization for the individual methods.
Rather, we choose a very exhaustive basis to numerically support the claims on analyticity 
and stability made above. For all calculations we chose a basis of standard polynomial finite
elements on $F$. Outside, we use a combination of polynomial
finite elements with polynomials times a decaying exponential $Q_n(x)\exp(\mp\al x)$ 
for the first and last element with $x\to\mp\infty$, respectively. 
A detailed description of this basis, 
which defines the ``infinite range''ECS method, can be found in \cite{scrinzi10:irecs}. 
Here we used $\al=6$ and polynomial orders $n=20$ and  $N\times N$ sizes of 
$N\approx 360 \sim 560$ for $\mS_\eta$ and $\mD_\eta$ matrices.
Clearly, with some optimization, significantly smaller bases can be used. However, 
as we wanted to present results independent of discretization issues, this significantly 
over-saturated basis was used.

Our initial state is 
\bel{initial}
\psi[0](x)=\left\{\begin{array}{cl} 
\cos^2(\pi x/4)e^{ix}&{\rm for}\quad x\in[-2,2]\\
0&{\rm else.}
\end{array}\right.
\eel

We choose $F=[-2.5,2.5]$ and propagate until $t=2.5$. At that time, the center of the 
initial distribution falls onto the right hand boundary $\partial F=2.5$. The accuracy of 
$\psi_\eta[t]$ is studied by comparison with a solution $\phi_\eta[t]$ on an extended domain
$F_\phi=[-5,5]$.

\subsection{The wave equation}

\begin{figure}[t]
  \centering \includegraphics[height=0.9\textwidth,angle=-90]{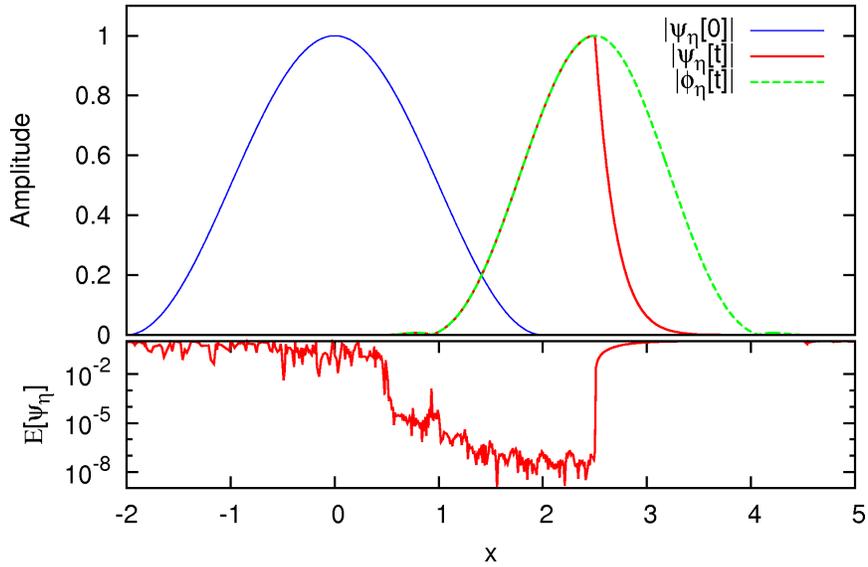}
  \caption{\label{fig:wave}(color online)
Relative error $E(x)$ of $\psi_\eta[t]$ at time $t=2.5$ 
for the wave equation computed with PML absorption. The error is defined in Eq.~(\ref{eq:error}).}
\end{figure}

Figure~\ref{fig:wave} shows our best PML solutions 
for the wave equation \ref{eq:wave} with initial state (\ref{eq:initial}) $\psi_\eta\up{PML}[t]$ and 
$\phi_\eta\up{PML}[t]$ at $t=2.5$. In the calculations, we used $\eta\up{PML}=5i$, but the results
are independent of the exact choice of $\eta$.  The relative error 
\bel{error}
E(x)=\frac{|\psi_\eta\up{PML}[t](x)-\phi_\eta\up{PML}[t](x)|}{|\psi_\eta\up{PML}[t](x)+\phi_\eta\up{PML}[t](x)|}
\eel
nowhere on $[0.5,2.5]$ exceeds $10^{-5}$ and near the edge it is $\sim 10^{-7}$. The remaining error
is largely do to numerical loss in our propagation scheme. 

Not surprisingly, PML works as expected
for the simple 1d wave equation.
For the reasons discussed above, the wave equation diverges 
when using ECS and no numerical results were obtained.

\subsection{The TDSE}
\begin{figure}[!ht]
  \centering \includegraphics[height=0.9\textwidth,angle=-90]{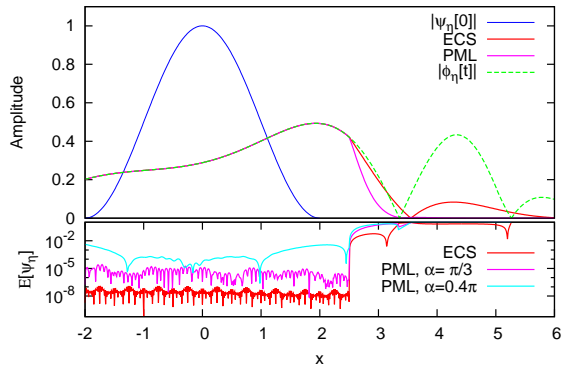}
  \caption{\label{fig:tdse}(color online)
Relative errors $E(x)$  of $\psi_\eta[t]$ at time $t=2.5$ for the TDSE computed 
with ECS and PML absorption, respectively. The error is defined in Eq.~(\ref{eq:error}).}
\end{figure}

Figure~\ref{fig:tdse} shows the best achieved ECS and PML solutions of the TDSE and their relative errors.
We used scaling parameters $\eta\up{ECS}=e^{i\th}=e^{i/2}$ and $\eta\up{PML}=5e^{i\al}, 
\al=\pi/8\sim\pi/3$ for ECS and
PML, respectively. 
Both methods perform well, where ECS is near the accuracy $\sim 10^{-8}$ achievable in our
numerical scheme. The error of the PML is constant, but about two orders of magnitude 
larger using the same, essentially saturated discretization. Errors, also for PML,  are very 
likely due to accumulated roundoff, as with the extremely dense discretization, both schemes 
lead to very small time steps. For $\eta\up{PML}$ with complex arguments $\al\gtrsim~0.4\pi$
PML deteriorates, as small frequencies cease to be damped (cf. Fig.~\ref{fig:tdse}).
The somewhat better numerical behavior of ECS may be due to 
the fact that the discretization used was originally designed for ECS and no attempt 
was made to find a more PML-specific discretization. 

It must be mentioned, though, that we could obtain the very good results only with the 
exponentially decaying basis at the end intervals. With a standard finite element basis 
on a finite box we were not able to converge the PML to better than a few percent. 
Our use of constant stretching by $\Th_F$,
rather than a asymptotically growing $\Th_F$, which is customary in PML, may be the reason for this
problem. ECS,
as was shown in \cite{scrinzi10:irecs}, does converge for the TDSE using a finite box, although
significantly larger absorption ranges are needed.
\section{Conclusion} 

In this work we have introduced a new unified formulation of the ECS and PML schemes
based on spectral analysis. 
Both methods are mathematically well-defined and heavily rely on analytical continuation.
This does not require the solutions to be analytic as functions
of the spatial coordinates. In fact, in the present formulation, we admit and explicitly 
use discontinuities in the $x$-dependence. For transforming to localized representations that 
exponentially decay at large distances, both methods require that the solutions 
of the original problem  have purely ``outgoing'' asymptotics. ECS justifiably is 
called ``complex stretching'' and ensues a rotation of the spectrum into 
the complex plane. PML, in contrast, at least for the 1d example treated here, 
is a ``complex gauge transform'' and corresponds to a uniform shift of spectral 
eigenvalues into the lower half complex plane. 

Our numerical studies show that the presented mathematical reasoning leads
to accurate and stable numerical schemes.
We believe that, apart from the specific analysis performed here, the approach based on 
spectral analysis and analytical continuation of unitary maps will be useful in designing 
absorption methods also for other wave equations. From a rigorous point of view,
the analysis is clearly limited to linear operators.
When the equation or a class of solutions can be considered linear at least 
in some asymptotic sense, the analysis readily provides a first estimate of spectral 
properties --- stability of time-evolution --- and asymptotic form --- 
spatially finite approximations of the solution. It is a simple tool 
to construct the explicit shape of the differential operators and to clarify 
possible domain questions.    

\section*{Acknowledgment}
We acknowledge support by the excellence cluster "Munich Center for Advanced
Photonics (MAP)", by the Austrian Science Foundation (FWF) project
"ViCoM" (FWF No F41) and by the ANR-FWF project "LODIQUAS" (FWF No
I830-N13).


 \end{document}